# Preliminary Results from a U.S. Demographic Analysis of SMiSh Susceptibility *

Cori Faklaris**, Heather Richter Lipford, and Sarah Tabassum, University of North Carolina at Charlotte

{cfaklari, richter, stabass2} @charlotte.edu

*in submission to CHI 2024   ** corresponding author

As adoption of mobile phones has skyrocketed, so have scams involving them. The text method is called "SMiShing," (aka "SMShing", or "smishing") in which a fraudster sends a phishing link via Short Message Service (SMS) text to a phone. However, no data exists on who is most vulnerable to SMiShing. Prior work in phishing (its e-mail cousin) indicates that this is likely to vary by demographic and contextual factors. In our study, we collect this data from N=1007 U.S. adult mobile phone users. Younger people and college students emerge in this sample as the most vulnerable. Participants struggled to correctly identify legitimate messages and were easily misled when they knew they had an account with the faked message entity. Counterintuitively, participants with higher levels of security training and awareness were less correct in rating possible SMiSH. We recommend next steps for researchers, regulators and telecom providers.

CCS CONCEPTS • **Security and privacy~Human and societal aspects of security and privacy~Social aspects of security and privacy** • Usability in security and privacy

## 1 INTRODUCTION

As adoption of mobile phones has skyrocketed [30], so have scams involving these devices [31]. By Q4 2022, the top contact method in U.S. Federal Trade Commission scam reports was the phone (20% text, 19% phone call) [27]. The text method is called "SMiShing," (aka "SMShing", or "smishing") in which a fraudster sends a phishing link via Short Message Service (SMS) text to a phone. Banks who partner in mobile payments networks are commonly impersonated, as are delivery companies, retailers, and communication providers [32]. However, no data exists on who is most vulnerable to SMiShing. Prior work in phishing (its e-mail cousin) [5,8,9,24] and our informal interviews with industry researchers indicates that SMiShing vulnerability is likely to vary by both demographic and contextual factors. This data is needed to identify how to best intervene to reduce and mitigate SMiShing, such as providing evidence for U.S. telecom providers to use optimal filters for SMiSh and to provide in-context warnings to mobile phone users.

Multiple studies will be needed to fully investigate this problem. As a first step, we have conducted and analyzed data from a large-scale survey of U.S. adult mobile phone users. In this paper, we answer the following research questions:

- RQ1: How many adult mobile phone users can correctly rate three random text messages as either Legitimate or Fraudulent, as determined with data from a U.S.-representative survey panel?
- RQ2: Which U.S. demographic groups are most vulnerable to SMiShing, as determined through statistical analysis of online survey ratings of text messages and selected responses to the messages?
- RQ3: To what extent is the vulnerability identified in RQ1a significantly associated with a lack of prior training or other relevant experiences?

To answer these questions, we designed an online assessment of people's ability to identify whether a simulated text message was "real" or "fake." We also collected a number of demographic and security-related cognitive and behavioral variables. We deployed the survey to a Qualtrics panel of U.S. adult mobile-phone users from June 26 to July 1, 2023. After data cleaning, we analyzed the resulting N=1,007 responses.

Overall, we found that participants had more difficulty in correctly identifying legitimate text messages than fraudulent ones. Troublingly, they overwhelmingly and significantly fell for SMiSH if the message entity was one that they thought they would have an account with. These results suggests that, in participants' minds, thinking that they had an account with the entity named in the message overrode all caution gleaned from prior experience or training, or effort to examine the source identifiers for clues as to whether the given text message was a SMiSH attack or legitimate text message. Further, and controlling for account knowledge: we found that participants scored significantly lower on our SMiShing assessment in younger age brackets and if they reported currently studying for a four-year degree; and scored significantly higher if they reported holding a job in the Educational Instruction and Library category. This suggests that, more broadly, younger people and those in school are most vulnerable to SMiSh, while people whose jobs denote a non-security expertise in judging information sources and credibility are among the least vulnerable.

Finally, we found that a low score on our SMiShing assessment was significantly more likely from a participant who reported frequently experiencing security breaches, and – more counterintuitively – from those receiving a greater-than-average amount of security training, or taking greater-than-average care to keep alert for phishing and other scams online. Taking the results as a whole, we suspect the existence of a "security expertise bias," in which participants who perceive themselves to be expert in staying alert for social engineering may be over-correcting and identifying too many messages as fraudulent and too few as legitimate, vs. those who with a non-security expertise in vetting information.

Based on these results, we recommend, first, that U.S. cellular and business regulators work with usability experts to design a verification system and trust indicators to highlight verified sources for SMS mobile messages. This would have the impact of making the SMS text system far more usable for consumers, as people of any expertise or skill would be easily able to see at a glance whether they could trust the source of a commercial message. Second, we recommend more research to determine whether a cause-and-effect relationship exists between high levels of security awareness training and vulnerability to SMiShing and what explains it, as this cross-sectional study can only determine what significantly accounts for variances in message ratings. Third, we see the need for developing more-nuanced messaging and education in how to perceive and judge information credibility, especially for young adults and college students.

In summary, our contributions are the following:

- Up-to-date knowledge of demographic susceptibility to scam messages for the era of mobile phones and widespread use of remote messaging.
- Examples of simulated "real" and "fake text messages and a survey protocol for use in research on SMiShing.
- Empirically based recommendations for researchers, regulators, and telecom providers.

## 2 RELATED WORK

Phishing and SMiShing are two types of cyberattacks that use social engineering techniques to trick users into revealing their personal or financial information via computer-mediated communication [12]. These attacks pose serious threats to the security and privacy of users, as well as the reputation and trustworthiness of organizations.

### 2.1 Phishing

Phishing is among the most common and well-studied forms of cyberattack [33]. Attackers use fraudulent emails to impersonate legitimate entities and solicit sensitive information from users [18]. Phishing emails often contain malicious links or attachments that lead users to fake websites or download malware onto their devices. These attacks can target individuals or organizations, and can have various motives, such as stealing money, identities, credentials, or intellectual property. Various methods have been researched to prevent or detect phishing attacks, which Hong summarized as "make things invisible" (ex: deploy machine learning on the back end to classify and filter away phish), develop better user interfaces, and provide effective training [18].

Several studies have investigated the factors that influence users' susceptibility to phishing attacks, such as the design of the email, the content of the message, the context of the situation, and the characteristics of the user [2,7–10,20,24]. Sheng et al. [24] were among the first to investigate demographic vulnerability. They designed a survey in which respondents were asked to play the part of "Pat Jones," an administrator for the fictional Baton Rouge University, and respond to four email messages, two of which were phishing and two of which were legitimate. In their N=1,001 sample, they found that women were more susceptible than men to phishing, and participants between the ages of 18 and 25 were more susceptible than other age groups, which they explained as due to differences in computer and web expertise among the groups. Their study also found that educational materials were effective in reducing participants' willingness to enter information into bogus webpages, but that they slightly decreased users' tendency to click on legitimate links.

Our study also employs the "Pat Jones" persona developed by Sheng et al. and displays simulated scam messages for participants to respond to. We find that younger people remain more vulnerable to social engineering attacks, but that the significant variance by gender has disappeared (Section 5.2.1).

**2.2 SMiShing**

SMiShing (aka "SMShing", or "smishing") is a relatively newer form of cyberattack that uses fraudulent SMS text messages to deceive users into clicking on malicious links or providing personal information. SMiShing messages often exploit users' emotions, such as fear, love, or greed, to induce them to take immediate action without verifying the source or the validity of the message. SMiShing attacks can also leverage users' trust in certain services or brands, such as banks, delivery companies, or online retailers [32], or even security or military authorities [28]. For example, a SMiShing attack on customers of the U.S. Fifth Third Bank led them to enter their credentials on a bogus website, thinking the bank had requested this to unlock their accounts [25]. An even bigger attack tricked customers of Czech Post into downloading a malicious app to their phones [3]. Recently, attackers have exploited COVID-19 information confusion and the global shift to remote messaging to motivate users with bogus messages from contact tracing websites, insurance, or vaccine providers [1]. SMiShing attacks are more difficult to detect than phishing attacks, as text messages have fewer cues and indicators than emails, such as sender's address, subject line, or spelling errors [34]. Furthermore, text messages are more likely to be read and responded to than emails, as they are perceived as more personal and urgent – leading marketers as well as scammers to send unsolicited texts to mobile numbers [6].

Relatively few researchers have systematically studied SMiShing vulnerability. An exception is Rahman et al., who conducted an experiment to randomly deliver two of four types of SMiSh (generic, personalized, spoofed, or voice-based, and with content for a variety of entities and using reward or fear motivations) to 10,000 participants. Of these, 28.7% responded to the messages, 15.8% clicked on malicious links, and 3.1% entered personal information into bogus webpages. The researchers found that the SMiShing attacks were more effective when they used personalized or spoofed messages, as they increased the perceived legitimacy and urgency for users to respond.

Our study draws on Rahman et al. for the attributes of our SMiSh content. We designed messages with similar entities, scenarios, source identifiers, user action asks, and motivations to click or respond. Our study contributes quantitative data for several variables that they identified as factors in SMiSh responses, such as urgency and curiosity. One difference is that we added items to discern the effect of participants knowing that they (their assigned persona, "yourself" or "Pat Jones") had an account with the message entity, which turned out to be a significant influence on correct responses. Because their study found that doctorates were disproportionately likely to fall for SMiSh, we added a question about whether participants had a doctorate. Our study found high rates of these doctorates rating messages incorrectly, but we discovered this score predictor to be non-significant when controlling for account knowledge (Section 5.2.1).

**3 METHOD**

To pursue answer to our research questions, we designed and deployed a web-based questionnaire programmed in Qualtrics to gather statistical data about which demographics are vulnerable to SMiShing.

**3.1 SMiShing Assessment Design**

First, we developed a series of simulated text messages, half based on legitimate real-world SMS messages and half based on real-world fraudulent SMS messages, to test how accurately participants could assess which are really a SMiSh message. We drew on prior work such as [23,24], along with actual SMS text messages provided by industry researchers or found in an internet search, in crafting each to include a URL, a type of entity likely to appear in either fraudulent or legitimate emails [23], mention of a reward motivator or a fear motivator to respond [23], and other "look and feel" clues to credibility [16] such as source identifiers, typos, and writing style. Each participant was randomly served three such messages out of 14 (7 fraudulent and 7 legitimate) and asked to rate them on a five-point scale: 1=Fraudulent, 2=Likely Fraudulent, 3=Not Sure, 4=Likely Legitimate, and 5=Legitimate. The messages are reproduced (Figure 1) and summarized (Table 1) below.

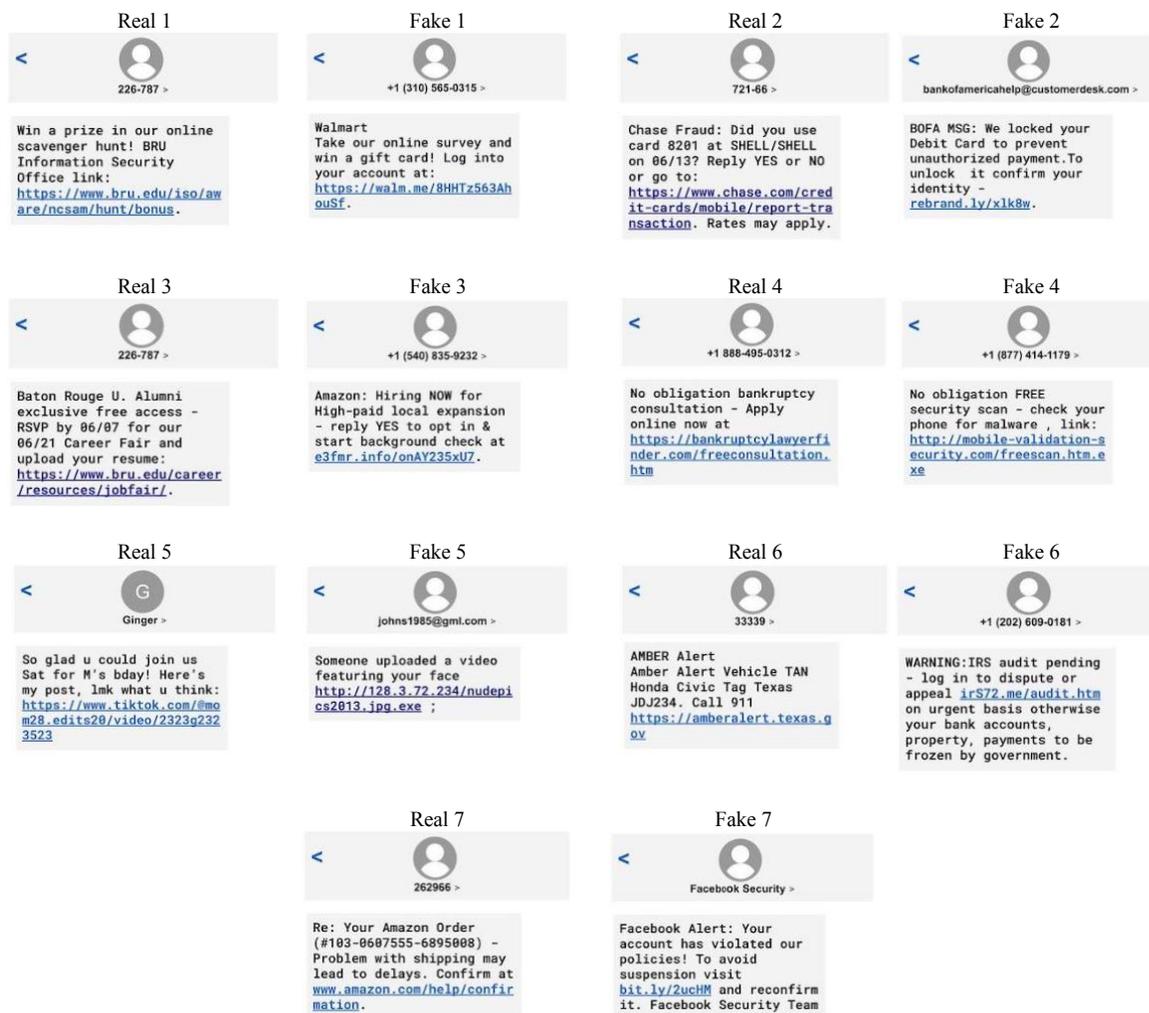

Figure 1: Three of these 14 simulated SMS text messages (7 "real" or legitimate, 7 "fake" or fraudulent) were randomly served to each participant to rate and answer questions about. Each included a URL and other clues to help guide the participants' judgments, such as whether it used an SMS short code or other source identifiers and whether it included typos or odd grammar and spacing.

Table 1: Summary of all 14 simulated messages' characteristics. The content of 3 implied a mix of reward and fear motivations, vs. 5 reward-motivated and 6 fear-motivated. All asked the user to click, with 3 asking also for a reply and 1 asking also for a phone call.

| ID | SMiSh? | Entity | Scenario | Source Identifiers | User Action Ask | Motivation |
|----|--------|--------|----------|--------------------|-----------------|------------|
| R1 | No | University | Web engagement | SMS short code, URL | Click | Reward |
| F1 | Yes | Retailer | Survey login | Phone number, URL | Click | Reward |
| R2 | No | Bank | Card fraud alert | SMS short code, URL | Click or Reply | Fear |
| F2 | Yes | Bank | Card fraud alert | Email address, URL | Click | Fear |
| R3 | No | University | Campus career fair | SMS short code, URL | Click | Reward |
| F3 | Yes | Retailer | Job offer | Phone number, URL | Click and Reply | Reward |
| R4 | No | Law office | Bankruptcy consult | Phone number, URL | Click | Reward/Fear |

| F4 | Yes | Security firm | Malware phone scan | Phone number, URL | Click | Reward/Fear |
| R5 | No | Phone contact | Private video | Contact name, URL | Click and Reply | Reward/Fear |
| F5 | Yes | Individual | Private video | Email address, URL | Click | Fear |
| R6 | No | Government | Missing child | SMS short code, URL | Click or Call | Reward |
| F6 | Yes | Government | Audit, seizure risk | Phone number, URL | Click | Fear |
| R7 | No | Retailer | Shipping delay | SMS short code, URL | Click | Fear |
| F7 | Yes | Social media | Account at risk | Contact name, URL | Click | Fear |

Next, we piloted our questionnaire with in-person cognitive or "think aloud" interviews [22,26] (*N*=2), review sessions with our study team (*N*=6) and remote surveys on Prolific (*N*=11). The most important piece of feedback we gathered was that knowing whether someone has an account with the entity in the SMS text message helps them judge whether it is fraudulent or legitimate. To address this feedback, we randomized all participants into two survey conditions: judging the SMS text messages as either "yourself" (described as whether you, the participant, had received the message on your phone) or as "Pat Jones" (adapted from [24], described as a staff member of Baton Rouge University who has many accounts and whose job makes it important to not fall for fraudulent text messages and to respond promptly to legitimate text messages). We also asked participants, at the end of each block of questions about an SMS text message, whether the entity mentioned was one that they had an account with, to be answered as "Yes," "No," or "Not Sure."

### 3.2 Data Collection

We hired Qualtrics to recruit a survey panel of at least 1,000 U.S. mobile phone users age 18 or older that roughly matched recent U.S. Census data for age, income, and education: 18-34: 30% / 35-54: 32% / 55+: 38%; <$50K: ~35% / $50K-100K: ~35% / 100K+: ~30%; no college degree: 65% / 4-year degree or higher: 35%. Participants who met these quotas were asked a series of other demographic questions: their more-specific brackets for age, income, and education; their gender and racial/ethnic identities; household size; experience with handling sensitive data; and occupation status and job category, per the U.S. Bureau of Labor Statistics. At the end of the survey, participants were asked an attention-check question and a series of items to assess their internet and information-security experiences, attitudes, behavior intentions, and prior training on how to respond to phish and SMiSh.

Along with demographics as identifiers, the questionnaire collected IP addresses and device metadata, to enable us to map responses and to test for effects from device modality and operating system. We did not collect other identifiers, to encourage our participants to answer freely and because other identification was not needed to answer our research questions. Our research design, recruitment and consent language, and survey and interview protocols were approved by our Institutional Review Board as an exempt study under Category 2 of the U.S. Revised Common Rule. See Appendix A.1 for the survey questions used in this study.

### 3.3 Procedure

Once the study team had reviewed the developed questionnaire and was satisfied with it, we provided Qualtrics with the URL to the coded online survey (Figure 2). Qualtrics then passed this link along to its third-party panel providers. Participants who clicked on the URL for the survey and click Yes for consent to participate were directed to a page to ask for their general demographic information and use of mobile phones. Those who checked boxes for demographic quotas that have already been reached, or who marked that they were under 18 or do not own a smartphone or feature phone, or who failed the CAPTCHA tests for fraud [13], were redirected to the Exit screen. This programming helped to ensure quality responses in the final dataset.

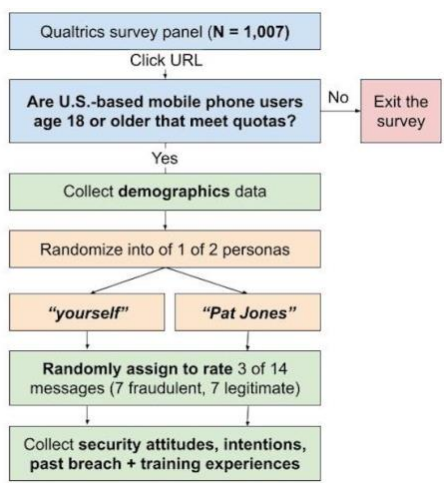

Figure 2: In our survey flow, participants who met the study qualifications were randomized into two conditions or "personas": either to rate text messages as "yourself" or as the "Pat Jones" persona. All were randomly served three of 14 text messages to rate and answer questions about. The survey also collected demographic information and data about people's security attitudes, their security behavior intentions, and their past experiences with security breaches and with training for security awareness and phishing / SMiShing mitigation.

The questionnaire accepted responses from June 26 to July 1, 2023. Once the quotas had been met, one member of the study team downloaded the responses to Microsoft Excel and conducted a visual inspection of the response patterns and typed answers to the open-ended text questions. Responses judged to be of bad quality responses were deleted. The master dataset was further cleaned, prepped for analysis, and uploaded to a secure, centralized cloud repository.

### 3.4 Participants

Qualtrics sourced responses from 1,000 people plus 1% overage. All had passed CAPTCHA fraud checks and answered affirmatively that they were U.S.-based internet users age 18 or older who owned a mobile phone and met our demographic quotas for age, education, and income. All had passed the attention-check item 2/3 of the way through the survey that directed them to answer with the 4[th] bullet point to retain their responses. The deletion of four responses with evidence of repeated nonsense copy-pastes into a text-input box resulted in a total dataset of N=1,007 (Tables 2-3).

Table 2: Counts for major demographic characteristics of participants. A little over a quarter were younger than age 25. About half were not currently in school and had attained less than a four-year college degree. The majority reported a household income of at least $50,000 per year, and most reported living in a household with other people. For gender, the majority identified as female. For race and ethnicity, the majority identified as being not Hispanic, Latino or Spanish, and as being White or Caucasian.

| Age | | Education | | Household Inc. | | Gender Identity | | Hisp./Lat./Sp.? | | Racial/Ethnic Identity | | Household Size | |
|---|---|---|---|---|---|---|---|---|---|---|---|---|---|
| 18-24 | 232 | No 4y deg. and not in school | 537 | < $26.5K poverty line | 188 | Female | 630 | No | 900 | White/Cauc. | 839 | 1 ppl. | 175 |
| 25-34 | 70 | No 4y deg., but in school | 129 | $26.5-$49K | 171 | Male | 362 | Yes | 96 | Black/African | 106 | 2 ppl. | 320 |
| 35-54 | 312 | | | $50-$99K | 358 | Nonbinary | 10 | Prefer not to say | 10 | Asian – total for all regions | 24 | 3 ppl | 160 |
| 55-64 | 173 | 4y deg., but no doctorate | 232 | $100K+ | 289 | Self-described | 2 | | | Native Am. or Alaska Native | 9 | 4 ppl. | 226 |
| 65+ | 219 | Has doctorate | 108 | | | Prefer not to say | 2 | | | Self-described | 15 | 5+ ppl. | 125 |
| | | | | | | | | | | Prefer not to say | 13 | | |

Table 3: Counts for participants' reported primary mobile phone and usage, experience working with sensitive data, and job characteristics. The vast majority used a smartphone primarily, and most reported using a mobile phone for more than 10 hours in the past week. About half had either "none at all" or "only a little" experience working with sensitive data. About half also reported working either full-time or part-time outside the home, with 2/3 of these in occupations that center on or heavily involve computers.

| Primary Mobile Phone | | Usage / Past Week | | Exp. w/Sens. Data | | Primary Job Status | | Top Occupations (FT or PT) | |
|---|---|---|---|---|---|---|---|---|---|
| Smartphone - Android | 531 | <6 hrs. | 167 | None at all | 390 | Full-time (FT) | 433 | Sales and Related | 51 |
| Smartphone - iOS/Apple | 452 | 6-10 hrs. | 208 | Only a little | 149 | Part-time (PT) | 112 | Business/Financial Ops. | 46 |
|  |  | 11-20 hrs. | 248 | A moderate amount | 191 | Unemployed | 148 | Computer/Mathematical | 44 |
|  |  | 21-30 hrs. | 190 |  |  | Retired | 134 | Office/Admin. Support | 44 |
| Other smartphone | 7 | >30 hrs. | 193 | A lot | 137 | At-home parent | 54 | Construction and Extraction | 38 |
| Featurephone with camera | 11 |  |  | A great deal | 139 | Self-employed | 42 | Healthcare Practitioner and Technical | 38 |
| Basic phone with no camera | 5 |  |  |  |  | Unable to work due to disability | 27 | Educational Instruction and Library | 37 |
|  |  |  |  |  |  | FT student | 43 | Food Prep. and Service | 36 |
|  |  |  |  |  |  | PT student | 13 | Management | 36 |

About 75% of participants reported receiving at least "a little" security awareness training, and about one-third reported receiving training specifically to help "identify fraudulent links or other threats in text messages." Further, a little more than half (51.5%) reported spotting and actively avoiding clicking on a suspected SMiSh message in the past three months. Other responses were "No, I have not noticed any fraudulent links in email, text messages, or web posts" (15.4%), "Yes, but it turned out to be a test being conducted as part of security awareness training" (7.0%), "Yes, and it turned out to be a scam, but nothing bad happened, to my knowledge" (15.7%), "Yes, and it turned out to be a scam, and I suffered a bad outcome (such as malware or theft of account credentials)" (4.4%), and Not Sure (6.1%).

### 3.5 Data Analysis

We calculated descriptive statistics and inferential statistics and drew figures using IBM SPSS and Microsoft Excel. The main tests used were one-way analyses of variance (ANOVAs), with post-hoc tests for pairwise comparisons, and multi-step linear and logistic regressions. The latter tests were used to assess the degree to which a predictor variable significantly accounted for variances in rating score and correctness, and to compare a control regression model with one that added a new predictor variable. We used model fit and a 95% confidence interval to assess statistical significance.

To score the SMiShing assessment described in Section 3.1, we counted as correct any answer for a simulated "fake" text message (F1-7) that was rated "Fraudulent" or "Likely Fraudulent," and any answer for a simulated "real" text message (R1-7) that was rated "Legitimate" or "Likely Legitimate." We used this scheme to compute a categorical variable, CORRECT, used as the outcome variable in logistic regressions; and a continuous variable, SCORE, used as the outcome variable in linear regressions. For CORRECT, we assigned a value between 3 and 0 depending on whether the participant had rated three, two, one, or zero messages correctly. For SCORE, we reverse-coded answers on the fraudulent text messages, then computed the average of the participant's ratings on the 1-5 scale of their assigned three messages, with a possible range of 1.00 (representing all being rated incorrectly and with confidence in that incorrectness) to 5.00 (representing perfect correctness and confidence in these correct ratings).

## 4 RESULTS

### 4.1 RQ1: Accuracy in Identifying Simulated SMiSh vs. 'Real' Text Messages

Overall, participants correctly identified whether the messages were legitimate or fraudulent 52.6% of the time, calculated by dividing the number of correct ratings (Likely Fraudulent or Fraudulent for the "fake" messages, or Likely Legitimate or Legitimate for the "real" ones) by total number of messages seen. We found that participants did much better at correctly identifying the simulated "fake" text messages (81.4%) than at correctly identifying the simulated "real" ones (23.5%) (Figures 3). Most participants reported receiving at least "a little" security training, which may have contributed to the high rates at which they could correctly spot the SMiSh. However, it may have also led them to over-correct and misidentify the legitimate messages, as happened in Sheng et al.'s study of phishing vulnerability and educational outcomes [24].

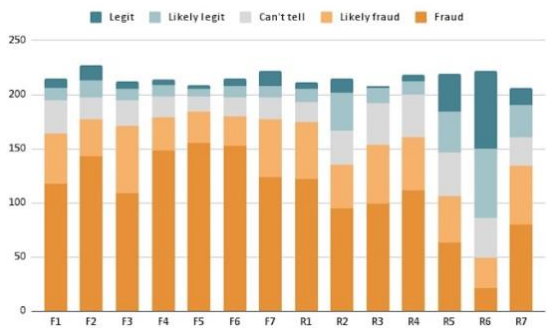

Figure 3: A majority of participants correctly rated all 7 simulated "fake" SMS text messages as Likely Fraudulent or Fraudulent. For 5 of 7 simulated "real" text messages, a majority of participants incorrectly rated them as Likely Fraudulent or Fraudulent.

Among the simulated "real" text messages, only R6 (the simulated "Amber Alert" text message) was correctly identified as Likely Legitimate or Legitimate by a majority (61.3%) of those who saw it, followed by R5 (with the Phone Contact identifier and link to a popular video platform) (32.9%). Among the simulated "fake" messages, participants did the best at correctly rating F5 (with the cryptic suggestion that the receiver's face was identifiable in nude images) as Likely Fraudulent or Fraudulent (88.0%). Participants tended to reply "Not Sure" more often for the simulated "real" text messages than for the simulated "fake" ones, suggesting that there were fewer interface indicators available to guide their judgments about legitimacy. Figure 4 shows a side-by-side comparison of messages, ordered by pairs of "real" and "fake" variations on similar entities, subjects, and/or motivations as described in Section 3.1, Figure 1 and Table 1.

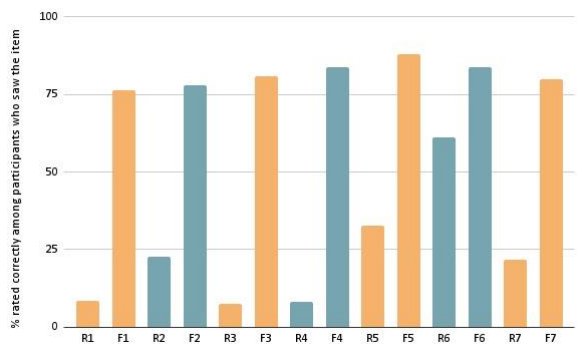

Figure 4: A side-by-side comparison of what percentage of participants who saw a given text message rated them correctly, ordered by pairs of "real" and "fake variations on similar entities, subjects, and/or motivations (Figure 1 and Table 1). A majority who saw the government-entity messages (R6, the "Amber Alert" message, and F6, the "tax audit and asset freeze" message) rated them correctly.

### 4.1.1 How and Why Participants Said They Would Respond to a Given Message

When asked how they would respond to any given simulated "fake" message (Figure 5), a minority of participants indicated they would report the message using device options such as clicking Block This Caller or Report Junk (38.7%), while a majority said they would delete it and/or ignore it (73.3%). Reponses were similar for the "real" messages (25.4% and 61.3%, respectively), which participants often incorrectly identified as SMiSh or likely SMiSh. While relatively few people selected "Reply to SMS text message to provide information" for the simulated "fake" messages (5.3%), some indicated that they would reply with STOP, BLOCK or other codes (17.5%). This still may accomplish the goal of the SMiSh attacker, since they may be testing the number to see if it remains in service and would be useful for a future scam [23]. Few participants said that they would respond in other ways that could meet the attacker's goals: click on the link (6.3%), forward the message to someone else (3.3%), or keep, save or archive the message (4.9). A minority said they would check the link on device, either by copy-pasting or typing the link into their phone's web browser (11.4%). Checking the link is a strategy recommended for phishing detection and mitigation [2,9,24], but is more easily accomplished on a larger device.

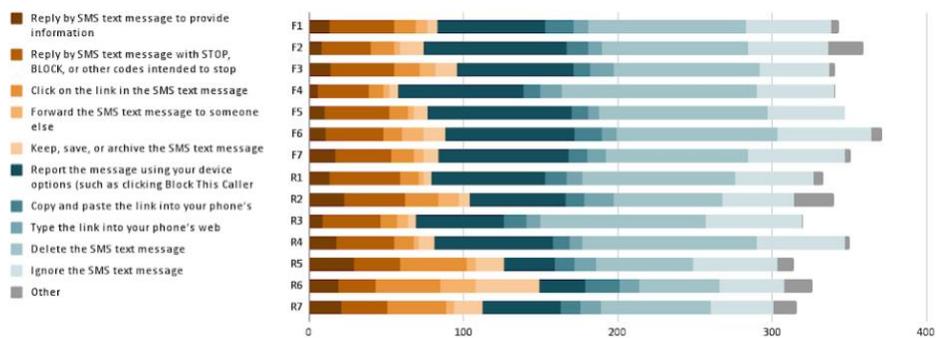

Figure 5: Counts of how many participants who saw a simulated text message indicated that they would respond with the given action. (The "Other" reasons are listed in Appendix A.2.) While relatively few people selected "Reply to SMS text message to provide information" for the simulated "fake" messages, a number indicated that they would reply with STOP, BLOCK or other codes. This still may accomplish the goal of the SMiSh attacker, since they may be testing the number to see if it remains in service and would be useful for a future scam.

When asked why they would respond a certain way (Figure 6), participants' responses were similar for the fraudulent messages as for the legitimate ones on four measures: sense of urgency (13.0% for "fake" vs. 14.2% for "real"), curiosity (10.9% for "fake" vs. 12.6% for "real"), seeking a good outcome for myself (12.3% for "fake" vs. 12.7% for "real"), and lack of interest in the message (36.4% for "fake" vs. 32.1% for "real"). Even for the legitimate messages, participants reported little trust in the sender (14.1%, versus 10.3% for the "fake" messages) or in the link URL (9.6%, versus 7.7% for the "fakes"), suggesting that these source indicators were only of marginal help in participants' assessments. Slightly more than half of participants who saw fraudulent messages reported "seeking to avoid a bad outcome for myself" as reasons for their response (50.7%), though a significant minority also reported this for the legitimate messages (47.0%).

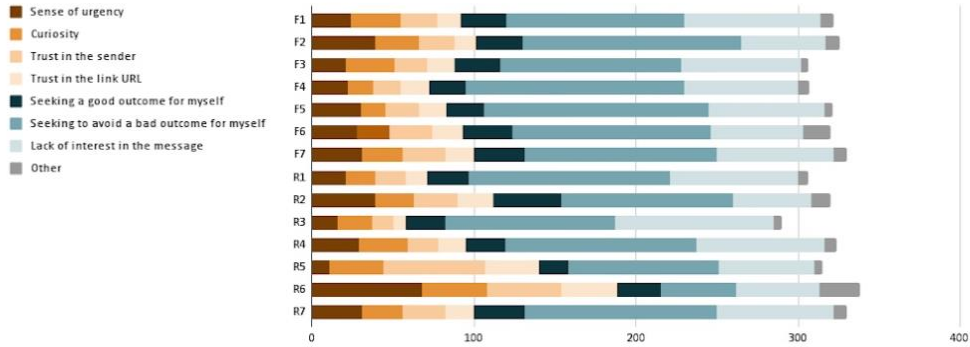

Figure 6: Counts of the reasons that participants selected their given responses to each simulated text message (Figure 5). Of the fraudulent messages, F3 (a fake retailer job offer) and F4 (a fake "security scam" to download malware) succeeded the most at inspiring a sense of urgency to respond. (The "Other" reasons are listed in Appendix A.2.)

*4.1.2 Influence of Persona Assignment or Account Knowledge on Accuracy of Ratings*

Next, to delve more deeply into the above results, we conducted tests to assess whether two variables that we theorized would influence people's correctness ratings– persona assignment and self-reported account knowledge – had statistically significant effects. Most tests revealed no significant difference in correctness depending on whether the participants had answered the questions as "yourself" or as "Pat Jones." We did, however, find a significant difference in CORRECT values by whether participants had answered Yes to a question asking them whether they knew that they had an account with the named entity (Table 4). Those in the "Pat Jones" condition were slightly yet significantly more likely to answer Yes to this question than those in the "yourself" group: t(1004)= -2.859, *p*=.004. This certainty of knowledge seemed to explain many cases where participants correctly identified the simulated "real" messages as Legitimate or Likely Legitimate, and is consistent with data collected during the survey pilots (Section 3.1). More concerning was that participants who answered Yes to the accounts-knowledge question were significantly *less* likely to rate a *"fake"* message correctly as Fraudulent or Likely Fraudulent. These results suggests that, in participants' minds, thinking that they had an account with the entity overrode all caution gleaned from prior experience or training, or effort to examine the source identifiers for clues as to whether the given text message was a SMiSH attack or legitimate text message.

Table 4: Results for logistic regression models estimating how likely a participant was to have correctly rated the given message, with the first added predictor being that they said Yes to a question asking whether they thought that they had an account with the given entity (Model 1), and the second added predictor being whether they were told to answer as "yourself" vs. "Pat Jones" (Model 2). A positive *b* coefficient is associated with higher-scoring participants being part of the reference group (Yes to account knowledge, or the "yourself" persona), while a negative *b* is associated with higher-scoring participants being in the non-reference group (No/Not Sure for account knowledge, or the "Pat Jones" persona). The Wald statistic tests for a significant difference in *b* from 0 at the <.05 level (bolded). The odds ratio shows whether the reference group is more (>1.00) or less (<.1.00) likely to have scored the message correctly, with a ratio above 3.000 or below 0.333 denoting a strong relationship between the predictor and the correct score [17]. This ratio is significant if the 95% CI does not include 1.000 (α =.05, bolded). Finally, a lower -2 log likelihood statistic indicates better model fit.

| ID | Model | Added predictor | *b* | SE | Wald | df | *p* | Odds ratio (exp(*β*)) | 95% CI Lower | 95% CI Upper | -2 log likelihood |
|---|---|---|---|---|---|---|---|---|---|---|---|
| R1 | 1 | Acct. Know. | **3.136** | 0.58 | 29.198 | 1 | **<.001** | **23.002** | **7.376** | **71.729** | 89.833 |
|  | 2 | Persona | -0.352 | 0.566 | 0.387 | 1 | 0.534 | 0.703 | 0.232 | 2.132 | 89.443 |
| F1 | 1 | Acct. Know. | **-1.730** | 0.343 | 25.419 | 1 | **<.001** | **0.177** | **0.090** | **0.347** | 212.509 |
|  | 2 | Persona | -0.111 | 0.344 | 0.105 | 1 | 0.746 | 0.895 | 0.456 | 1.757 | 212.404 |
| R2 | 1 | Acct. Know. | **2.843** | 0.431 | 43.443 | 1 | **<.001** | **17.164** | **7.370** | **39.97** | 170.169 |
|  | 2 | Persona | -0.116 | 0.428 | 0.074 | 1 | 0.786 | 0.89 | 0.384 | 2.061 | 170.095 |
| F2 | 1 | Acct. Know. | **-1.787** | 0.349 | 26.19 | 1 | **<.001** | **0.167** | **0.084** | **0.332** | 212.602 |
|  | 2 | Persona | -0.193 | 0.344 | 0.313 | 1 | 0.576 | 0.825 | 0.42 | 1.619 | 212.289 |

| ID | Model | Added predictor | b | SE | Wald | df | p | Odds ratio (exp($\beta$)) | 95% CI Lower | 95% CI Upper | -2 log likelihood |
|---|---|---|---|---|---|---|---|---|---|---|---|
| R3 | 1 | Acct. Know. | **1.363** | 0.562 | 5.889 | 1 | **0.015** | **3.906** | 1.300 | 11.742 | 107.474 |
|  | 2 | Persona | 0.453 | 0.545 | 0.690 | 1 | 0.406 | 1.573 | 0.540 | 4.577 | 106.768 |
| F3 | 1 | Acct. Know. | **-1.157** | 0.37 | 9.801 | 1 | **0.002** | **0.314** | 0.152 | 0.649 | 198.832 |
|  | 2 | Persona | 0.119 | 0.36 | 0.109 | 1 | 0.742 | 1.126 | 0.556 | 2.282 | 198.723 |
| R4 | 1 | Acct. Know. | **2.618** | 0.555 | 22.238 | 1 | **<.001** | **13.715** | 4.619 | 40.721 | 101.584 |
|  | 2 | Persona | **1.378** | 0.591 | 5.429 | 1 | **0.020** | **3.967** | 1.245 | 12.643 | 95.503 |
| F4 | 1 | Acct. Know. | **-1.970** | 0.427 | 21.264 | 1 | **<.001** | **0.139** | 0.060 | 0.322 | 170.284 |
|  | 2 | Persona | 0.048 | 0.395 | 0.015 | 1 | 0.903 | 1.049 | 0.484 | 2.277 | 170.269 |
| R5 | 1 | Acct. Know. | **2.087** | 0.326 | 41.058 | 1 | **<.001** | **8.063** | 4.258 | 15.268 | 232.037 |
|  | 2 | Persona | -0.228 | 0.325 | 0.491 | 1 | 0.484 | 0.796 | 0.421 | 1.506 | 231.544 |
| F5 | 1 | Acct. Know. | **-2.924** | 0.495 | 34.900 | 1 | **<.001** | **0.054** | 0.020 | 0.142 | 114.696 |
|  | 2 | Persona | 0.030 | 0.493 | 0.004 | 1 | 0.952 | 1.030 | 0.392 | 2.710 | 114.692 |
| R6 | 1 | Acct. Know. | 0.627 | 0.322 | 3.796 | 1 | 0.051 | 1.873 | 0.996 | 3.52 | 292.464 |
|  | 2 | Persona | 0.171 | 0.278 | 0.378 | 1 | 0.539 | 1.187 | 0.688 | 2.047 | 292.086 |
| F6 | 1 | Acct. Know. | **-2.143** | 0.413 | 26.965 | 1 | **<.001** | **0.117** | 0.052 | 0.263 | 161.237 |
|  | 2 | Persona | -0.412 | 0.421 | 0.958 | 1 | 0.328 | 0.662 | 0.290 | 1.511 | 160.268 |
| R7 | 1 | Acct. Know. | **2.440** | 0.468 | 27.135 | 1 | **<.001** | **11.476** | 4.582 | 28.744 | 177.183 |
|  | 2 | Persona | 0.163 | 0.373 | 0.191 | 1 | 0.662 | 1.177 | 0.567 | 2.445 | 176.992 |
| F7 | 1 | Acct. Know. | **-1.917** | 0.371 | 26.742 | 1 | **<.001** | **0.147** | 0.071 | 0.304 | 193.778 |
|  | 2 | Persona | -0.252 | 0.362 | 0.486 | 1 | 0.486 | 0.777 | 0.383 | 1.579 | 193.290 |

### 4.2 RQ2 and RQ3: Differences in Message Scores Among Comparison Groups

*4.2.1 RQ2: SMiShing Vulnerability by Demographics*

Controlling for account knowledge, we found that variances in participants' message scores could be significantly explained by their different age brackets, by whether they reported currently studying for a four-year degree, and by whether they reported holding a job in the Educational Instruction and Library category (Table 5). The effect of having a doctorate, which was discovered to be significantly associated with falling for SMiSh in a prior study's sample [23], just missed being significant at the p<.05 level in our study once account knowledge was controlled for, as did age younger than 35, general employment status and the Office/Administrative job category. We found no significant effects on SCORE at the *p*<.20 level when controlling for account knowledge for the following demographics: income level, gender identity, Hispanic/Latinx/Spanish identity, other racial or ethnic identity, household size, or mobile phone type or usage.

Table 5: Selected demographic SCORE predictors in linear regression models, using account knowledge as a control predictor at a prior step. The Δ R² column shows the additional variance in SCORE added by the predictor vs. the control model. The standardized parameter estimate (β) shows the strength and direction of the predictor's effect on SCORE. Predictors with test statistics significant at the p<.05 level are and bolded. Control statistics are omitted for brevity.

| Predictor | Δ R² vs. control | Unstandardized est. b | SE | Standardized estimate (β) | t | p | 95% CI Lower | 95% CI Upper |
|---|---|---|---|---|---|---|---|---|
| **AGE ALL** | **0.004** | 0.038 | 0.019 | **0.065** | 1.998 | **0.046** | 0.001 | 0.075 |
| 18-34 | 0.004 | -0.113 | 0.058 | -0.062 | -1.941 | 0.053 | -0.226 | 0.001 |
| 55 or older | 0.002 | 0.074 | 0.056 | 0.043 | 1.319 | 0.187 | -0.036 | 0.183 |
| EDU ALL | 0.000 | 0.006 | 0.025 | 0.008 | 0.261 | 0.795 | -0.042 | 0.055 |
| **In school** | **0.011** | 0.225 | 0.084 | **0.105** | 2.671 | **0.008** | 0.060 | 0.391 |
| Doctorate | 0.009 | 0.180 | 0.099 | 0.103 | 1.812 | 0.071 | -0.015 | 0.375 |

| Predictor | Δ R² vs. control | Unstandardized est. b | SE | Standardized estimate (β) | t | p | 95% CI Lower | Upper |
|---|---|---|---|---|---|---|---|---|
| EMPL ALL | 0.003 | 0.026 | 0.014 | 0.058 | 1.835 | 0.067 | -0.002 | 0.053 |
| Office/Admin. | 0.004 | 0.248 | 0.128 | 0.061 | 1.935 | 0.053 | -0.004 | 0.500 |
| **Educ./Library** | **0.007** | 0.360 | 0.139 | **0.081** | 2.586 | **0.010** | 0.087 | 0.634 |
| Retired | 0.003 | 0.131 | 0.078 | 0.053 | 1.686 | 0.092 | -0.021 | 0.284 |
| Unemployed | 0.002 | -0.116 | 0.075 | -0.049 | -1.547 | 0.122 | -0.263 | 0.031 |

Using a Bonferroni correction to adjust *p* values for increase in Type I error risk, we further compared subgroups of our demographic variables to test for statistically significant differences in SCORE. We found no significant pairwise comparisons by job category or employment status. While we found no pairwise comparisons by age bracket that were significant at the adjusted *p* value, mean SCORE values show a clear positive association with an overall increase in age (Figure 7).

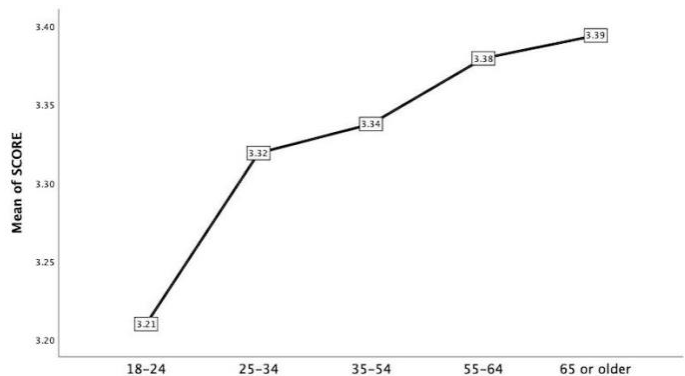

Figure 7: The line chart shows that participants' mean SCORE values increased in each successive age bracket. While no pairwise comparisons are significant at a Bonferroni-adjusted *p* value, linear regression found these differences to be significant overall when controlling for account knowledge (Table 5). It suggests that younger people may be more vulnerable to SMiShing attacks.

We did find two pairs of pairwise comparisons that were significant at the Bonferroni-adjusted p value: between those In School for a 4-year Degree vs. Not In School, No 4-year Degree; and those In School for a 4-year Degree vs. 4-year Degree and No Doctorate (Figure 8). In fact, mean SCORE values for participants who were In School for a 4-year Degree had a *negative* association with Account Knowledge (Table 6). It adds evidence to our theory that, in some participants' minds, thinking that they had an account with the entity overrode all other considerations as to whether the message was a SMiSH attack or legitimate text message (Section 4.1.2).

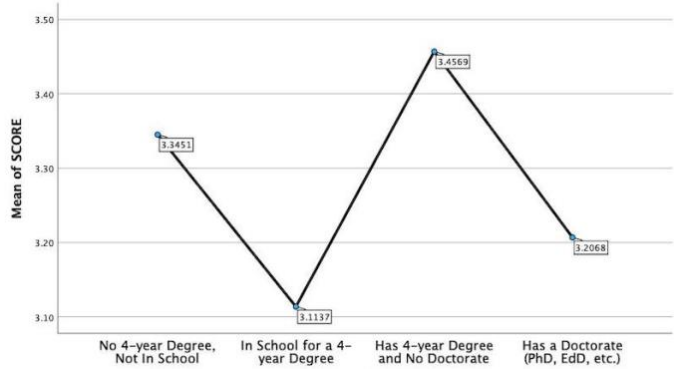

Figure 8: Mean SCORE values for those who are In School for a 4-year Degree were significantly lower (as shown by a Bonferroni-adjusted *p* value) than mean SCORE values for those who are Not In School and have No 4-year Degree, and from those who have a 4-year Degree and No Doctorate. This suggests that college students are a demographic group that is vulnerable to SMiShing.

Table 6: For those reporting being In School for a 4-Year Degree, mean SCORE values decreased for each message entity that they thought they had an account with. This adds evidence that, in some participants' minds, thinking that they had an account with the entity in the simulated text message overrode all other considerations in judging it to be legitimate or fraudulent (Section 4.1.1).

| Educ. Level | Acct. Know. | SCORE Mean | SE | 95% CI Lower | Upper |
|---|---|---|---|---|---|
| Not In School and Has No 4-year Degree | 0.00 | 3.283 | 0.045 | 3.194 | 3.373 |
| | 0.33 | 3.507 | 0.075 | 3.361 | 3.654 |
| | 0.67 | 3.307 | 0.117 | 3.078 | 3.536 |
| | 1.00 | 3.380 | 0.154 | 3.078 | 3.681 |
| **In School for a 4-year Degree** | **0.00** | **3.271** | 0.112 | 3.052 | 3.490 |
| | **0.33** | **3.171** | 0.136 | 2.904 | 3.437 |
| | **0.67** | **3.086** | 0.242 | 2.612 | 3.560 |
| | **1.00** | **2.709** | 0.166 | 2.383 | 3.035 |
| Has 4-year Degree and No Doctorate | 0.00 | 3.471 | 0.073 | 3.329 | 3.614 |
| | 0.33 | 3.564 | 0.109 | 3.349 | 3.779 |
| | 0.67 | 3.283 | 0.174 | 2.943 | 3.624 |
| | 1.00 | 3.161 | 0.208 | 2.753 | 3.569 |
| Has a Doctorate (PhD, EdD, etc.) | 0.00 | 3.329 | 0.147 | 3.041 | 3.618 |
| | 0.33 | 3.170 | 0.158 | 2.861 | 3.479 |
| | 0.67 | 3.300 | 0.226 | 2.857 | 3.743 |
| | 1.00 | 3.061 | 0.145 | 2.776 | 3.347 |

*4.2.2 RQ2: SMiShing Vulnerability by Security-Relevant Training or Experiences*

Controlling for account knowledge, we found that variances in participants' message scores were significantly associated with the frequency of their personal experiences of security breaches, the amount of their security awareness training, and their scores on the Security Behavior Intentions Scale (SeBIS) [11] subscale for Proactive Awareness (Table 7). The association of SCORE with all three security-relevant variables was negative – in other words, a low SCORE on the SMiShing assessment was significantly *more* likely from a participant who reported frequently experiencing security breaches, receiving a greater-than-average amount of security training, or taking greater-than-average care to keep alert for phishing and other scams online. We found no significant associations with SCORE at the *p*<.20 level when controlling for account knowledge for the following variables: frequency of hearing or seeing news about security breaches, amount of experience working with sensitive data, whether they reported clicking on SMiSh in the past three months, whether they specifically had received training on spotting and dealing with fraudulent text messages (included in Table 7), and scores on the Social Strategy subscale of the recently published Smartphone Security Behavior Scale (SSBS) [19].

Table 7: Selected security-relevant SCORE predictors in linear regression models, using account knowledge as a control predictor at a prior step. The **Δ R²** column shows the additional variance in SCORE added by the predictor vs. the control model. The standardized parameter estimate (β) shows the strength and direction of the predictor's effect on SCORE. Predictors with test statistics significant at the p<.05 level are and bolded. Control statistics are omitted for brevity.

| Predictor | Δ R² vs. control | Unstandardized est. b | SE | Standardized estimate (β) | t | p | 95% CI Lower | Upper |
|---|---|---|---|---|---|---|---|---|
| Breach-Personal | **0.007** | -0.064 | 0.024 | **-0.088** | -2.712 | **0.007** | -0.111 | -0.018 |
| Breach-Close tie | 0.002 | -0.037 | 0.023 | -0.050 | -1.582 | 0.114 | -0.082 | 0.009 |
| Security Training | **0.004** | -0.045 | 0.021 | **-0.069** | -2.115 | **0.035** | -0.086 | -0.003 |
| SMiSh Training | 0.000 | -0.022 | 0.058 | -0.012 | -0.372 | 0.710 | -0.136 | 0.092 |
| Security Attitude | 0.002 | -0.050 | 0.037 | -0.043 | -1.351 | 0.177 | -0.124 | 0.023 |
| SeBIS subscale | **0.024** | -0.147 | 0.03 | **-0.164** | -4.944 | **<.001** | -0.206 | -0.089 |

Using a Bonferroni correction to adjust *p* values for increase in Type I error risk, we further compared subgroups of our security-relevant variables to test for statistically significant differences in SCORE. While we found no pairwise comparisons by frequency of personal security breache experiences that were significant at the adjusted *p* value, mean SCORE values show a clear negative association with breach experience frequency (Figure 9).

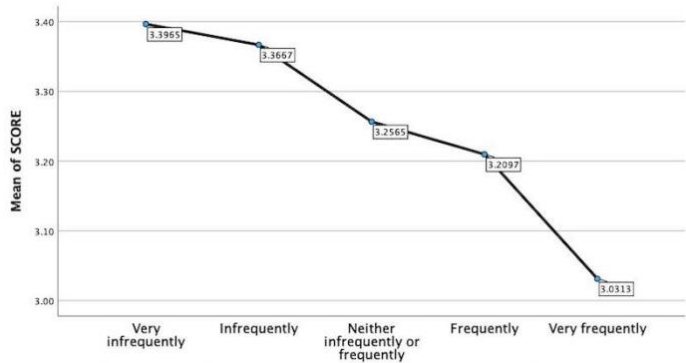

Figure 9: The line chart shows a negative association between participants' mean SCORE values and the frequency with which they reported experiencing security breaches. While no pairwise comparisons are significant at a Bonferroni-adjusted *p* value, linear regression found these differences to be significant overall when controlling for account knowledge (Table 7). It suggests that those who struggle with correctly distinguishing legitimate from scam text messages are more vulnerable to harms than others.

We found two pairwise comparisons on mean SCORE values in Figure 9 for amount of security awareness training that were significant at the Bonferroni-adjusted p value: between those with "None at all" vs. "A great deal" and those with "A moderate amount" vs. "A great deal" (Figure 10). While it is possible that trainees are taking away the wrong lessons from this instruction, it is also possible that those who are more vulnerable to social engineering attacks such as SMiSh do end up receiving more training – thus accounting for why those with "a great deal" of training scored significantly poorly.

Finally, we also compared the mean SeBIS subscale values (possible range: 1.00 to 5.00) according to how many messages each participant had correctly rated (possible range: 0 to 3). Three pairwise comparisons are significant at the Bonferroni-adjusted p value: between those who rated 0 and 2 correctly, between those who rated 0 and 3 correctly, and beween those who rated 1 and 3 correctly (Figure 11). Together with the finding on security awareness training and those at the top of Section 4, it suggests an "expertise bias" – that those who are expert in staying alert for social engineering may be over-correcting, identifying too many messages as fraudulent and too few as legitimate. However, more research is needed in these cases to determine whether a cause-and-effect relationship exists and what explains it, as this cross-sectional study can only determine what significantly accounts for variances in message ratings.

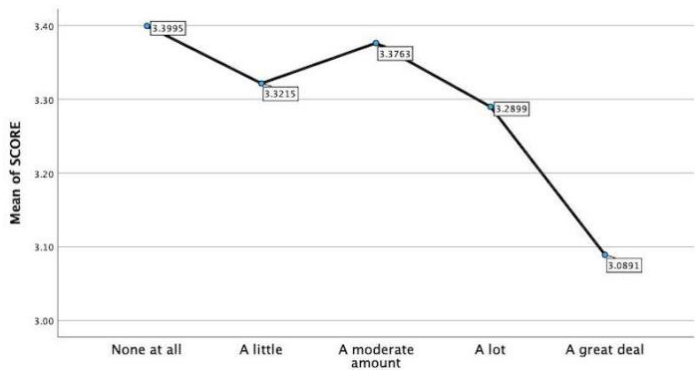

Figure 10: The line chart shows a negative association between participants' mean SCORE values (from 1.0000 to 5.0000) and the amount of formal security training that they have received on the job or in school. Linear regression found the overall differences to be significant when controlling for account knowledge (Table 7). More research is needed to determine the reason for this association.

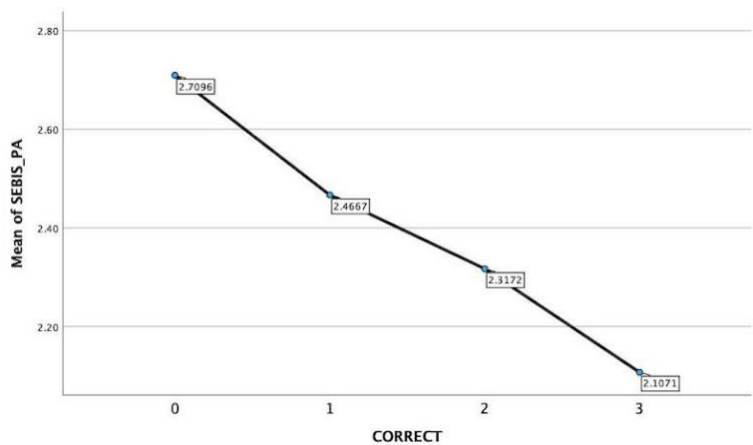

Figure 11: The line chart shows a negative association between participants' mean values on the SeBIS Proactive Awareness subscale (from 1.0000 to 5.0000) [11] and the number of text messages that they correctly rated. Linear regression found the overall differences to be significant when controlling for account knowledge (Table 7). More research is needed to determine the reason for this association.

## 5 DISCUSSION

Our results above show that, first, participants across all demographic groups struggled to correctly identify legitimate text messages regardless of source indicators that are available today in U.S. messaging interfaces, and that they fell for the simulated SMiSH if the message entity was one that they thought they would have an account with. Second, when drawing comparisons and controlling for account knowledge, we found that younger people and college students in our sample are the most vulnerable demographics for SMiShing attacks. Third, while significantly high scores for those with an Educational Instruction and Library job suggests that a non-security expertise in perceiving and judging information credibility is protective against SMiShing, we found evidence that a security expertise (as shown by high self-reported levels of security training and awareness) was associated with increased vulnerability to SMiShing.

Based on these results, we recommend, first, that U.S. cellular and business regulators work with usability experts to design a verification system and trust indicators to highlight verified sources for SMS mobile messages. This would have the impact of making the SMS text system far more usable for consumers, as people of any expertise or skill would be easily able to see at a glance whether they could trust the source of a commercial message. It could reduce the frequency with which scammers could trick mobile phone users by faking the name of a well-known entity with an easy-to-miss misspelling, such as "Amazom" or "Facebo0k Security." It could reduce the frequency of mobile phone users experiencing security breaches through reducing their susceptibility to falling for SMiSh. We discuss this in Section 5.1.

Second, we see the need for developing more-nuanced messaging and education in how to perceive and judge information credibility, especially for young adults and college students. However, our results also suggest that current security messaging and training is falling short of helping people strike the right balance in their judgments of SMiShing. So, we recommend more research to determine whether a cause-and-effect relationship exists between high levels of security awareness training and vulnerability to SMiShing and what explains it, as this cross-sectional study can only determine what significantly accounts for variances in message ratings. We discuss this in Section 5.2.

### 5.1 Helping Mobile Users with Identifying Legitimate Senders

Our first recommendation is that U.S. cellular and business regulators work with usability experts to design a verification system and trust indicators to highlight verified sources for SMS mobile messages. This would have the impact of making the SMS text system far more usable for consumers, as people of any expertise or skill would be easily able to see at a glance whether they could trust the source of a commercial message. It could reduce the frequency with which scammers could trick mobile phone users by faking the name of a well-known entity with an easy-to-miss misspelling, such as "Amazom" or "Facebo0k Security." It could reduce the frequency of mobile phone users experiencing security breaches through reducing their susceptibility to falling for SMiSh.

For such indicators to be reliable and trustworthy, they will need to be linked to a back-end verification system that cannot be easily "gamed" or hacked. (An example of how this can go wrong is the Twitter microblogging app's 2022 change to the "blue checkmark" verification rules, which enabled impersonation of the Eli Lilly & Co. branded account for a small payment and sparked a U.S. financial and political uproar [29].) There already exist some verification systems and indicators that governments and telecom providers around the world use to signal to mobile phone users that some messages should be trusted. In India, the Telecom Regulatory Authority of India (TRAI) has mandated the use of a special header for all bulk SMS messages sent by government agencies, banks, and other entities [35]. The header, and SMS short code, used in this header is different from the header and identifiers displayed in unverified messages (Figure 12). In Singapore and Australia, the governments use digital signatures that attach a cryptographic code to send secure and verified messages to citizens, through SingPass [36] and myGov Inbox [37], respectively. In South Korea, mobile network operators have been authorized to perform Pass identity verification in the form of challenge questions and responses through text messages [21]; in return, the mobile operators are allowed to collect and retain personal data.

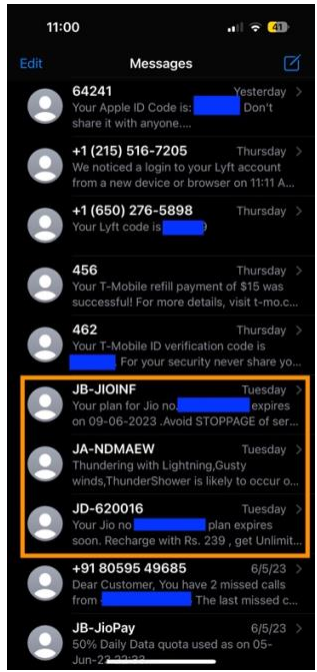

Figure 12: A screenshot of an Indian national's SMS inbox. The "XX-" prefixes indicate that TRAI has verified the message sender.

In the U.S., the Federal Communications Commission (FCC) has adopted a framework called STIR/SHAKEN that combats spoofed robocalls [38]. Using the framework, calls can be "signed" as legitimate and validated by the originating telecom providers, then digitally validated as the call is handed off among interconnected phone networks. We think this or a similar framework could be leveraged to mark some SMS text messages as coming from a verified sender, either with a special SMS header similar to India's, or a graphical visual cue, such as a green checkmark or star emoji.

### 5.2 Investigating Latent Factors and Improving Education and Training for SMiSh

Taking the results as a whole, we suspect the existence of a "security expertise bias," in which those who perceive themselves to be expert in staying alert for social engineering may be over-correcting and identifying too many messages as fraudulent and too few as legitimate, vs. those who with a general non-security expertise with vetting information. Sheng et al. documented similar outcomes when testing the effect of phishing interventions such as a comic strip and a quiz game [24]. With our results, one explanation is that the participants are coming away from security awareness training with either too-simplistic understandings of what signifies a threat or misunderstandings of how to judge a legitimate message (ex: taught to look for an entity that they know they have an account with, but not how to reason about whether the message is spoofing that entity, or without enough reinforcement that they remember to check for spoofing). However, it is also possible that those who are more vulnerable to social engineering attacks such as SMiSh simply end up receiving more training – thus accounting for why those with "a great deal" of training scored significantly poorly. We recommend that researchers conduct further studies to determine whether a cause-and-effect relationship exists and whether deficiencies exist in commonly used materials for security awareness training when it comes to SMiSh.

We also note that training is only one of the components of how people determine source credibility in contexts such as unsolicited text messages. Birnbaum and Stegner broke credibility or "believability" into three constructs: expertise, bias, and the person's point of view within an interaction [4], with expertise comprised further of training, experience and ability. Their experiments, in which participants were given various types of information to judge a used car's sale value, found evidence that expertise will magnify the effects of bias in how much weight people give to various information sources. Our finding that participants in Educational Instruction and Library jobs did significantly better at judging

SMiShing may point to this holistic view of credibility as useful in a SMiShing context. We theorize that such participants have had substantial amounts of training, experience, and ability in how to weigh source indicators and some amount of self-awareness about possible biases in their own thinking. We recommend that security educators explore methods to embed training within other contexts for boosting people's ability to correctly judge information, or that they examine this non-security instruction for constructs that are useful to boost the effectiveness of security training, or both.

Finally, we are alarmed to see that, in our study, younger people and those in school for a four-year degree were found to be significantly vulnerable to falling for SMiSh. We recommend that U.S. high schools, colleges, and universities either implement or adjust their training for information security so that students gain a more-sophisticated understanding of how to spot a fraudulent text message -- and what indicates that the text message is likely legitimate.

## 6  LIMITATIONS AND FUTURE WORK

Our survey provides useful statistical data for assessing how well U.S. participants were able to distinguish fraudulent from legitimate text messages. This cross-sectional design is not sufficient to establish cause-and-effect. In future work, we will recruit participants for in-depth interviews to get more context around how and why they rate the simulated text messages as being either fraudulent or legitimate; and how recently in time, and to what extent, they received security-relevant training. Our survey's results also suggest that users need help to identify legitimate messages more readily. In future work, we will test interface design improvements for mobile and wearable interfaces, such as indicators or a naming scheme for the SMS short codes, that can provide prominent cues to which SMS text messages are from legitimate sources. We practiced a careful method of iterative survey development to ensure its clarity and comprehensibility, and the anonymous method encouraged full honesty in answers. However, like all survey studies, ours is subject to a number of biases, such as self-report bias and social desirability bias, that possibly have skewed the results. A replication of this survey would help to validate the results and interpretations of this data. Finally, we developed a useful way to simulate SMiSh without sending participants unsolicited text messages that could have panicked them or led them to feel tricked once debriefed. In a future study, we may explore how to conduct a more true-to-life SMiSh test similar to Rahman et al. [23] that minimizes harms and boosts ecological validity.

## 7  CONCLUSIONS

In this study, we collected and analyzed data from a survey panel of N=1,007 U.S. adult mobile phone users. We found that younger people and college students were significantly vulnerable to SMiSh, that participants overall struggled to identify legitimate text messages, and that participants were easily misled if the fraudulent text messages mentioned an entity that they thought they had an account with. Our study contributes up-to-date knowledge of demographic susceptibility to scam messages for the era of mobile phones and widespread use of remote messaging, and examples of simulated "real" and "fake text messages and a survey protocol for use in research on SMiShing. Finally, we provide recommendations based in our data for use by researchers, regulators, and telecom providers. We hope these findings, and any future work based on it, will meaningfully improve the user experience and security of the U.S. mobile internet.

## ACKNOWLEDGMENTS

We are grateful to Carrie Gates and Guy V. Pearson of Bank of America and to Kaylei Goff of Winthrop University for their invaluable help with designing and carrying out this research, and to Jacqueline White for her feedback. This study was funded by the Center for Cybersecurity Analytics and Automation (https://www.ccaa-nsf.org/).

## A APPENDICES

### A.1 Survey Questions

*7.1.1 Consent Form – Study Summary Section*

You are invited to participate in a research study. Participation in this research study is voluntary. The information provided is to help you decide whether or not to participate. If you have any questions, please ask.

**Important Information You Need to Know**

- The purpose of this study is to document how different types of people interpret various mobile phone text messages.
- You must be a U.S. mobile phone user age 18 or older to participate in this study.
- You will be asked to answer a number of questions in an online survey. Most are multiple choice.
- If you choose to participate, it will require 15 minutes of your time.
- Risks or discomforts from this research are rare, but may include mild annoyance or frustration with the questions being asked, or discomfort with personal reflections about past experiences.
- While you will receive no direct benefit, the knowledge gained will be of value for improving people's experiences while using mobile phones and receiving text messages.
- If you choose not to participate, you may exit the survey.
- If you participate and complete the study, you will receive compensation in a form that is consistent with your agreements with the survey panel provider.

Please read this form and ask any questions you may have before you decide whether to participate in this study.

*7.1.2 Demographics*

**What is your age bracket?**
o  Under 18
o  18-34
o  35-54
o  55 or older

*Skip To: End of Block If What is your age bracket? = Under 18*

*Display This Question:*
*If What is your age bracket? = 18-34*

**Is your age 18 to 24?**
o  Yes
o  No

*Display This Question:*
*If What is your age bracket? = 55 or older*

**Is your age 65 or older?**
o   Yes
o   No

**What kind of mobile phone do you carry with you most often?**
o   A smartphone that can add apps, running Android (such as the Samsung Galaxy, OnePlus, or Google Pixel)
o   A smartphone that can add apps, running iOS (such as the Apple iPhone)
o   Another type of smartphone that can add apps, but not running Android or iOS (such as BlackBerry or Windows Mobile)
o   A feature phone that cannot add apps but has a camera (such as Kyocera, Jitterbug, or Nokia flip phones)
o   A basic mobile phone that has no camera (such as Easyfone or Punkt)
o   I do not use a mobile phone

*Skip To: End of Block If What kind of mobile phone do you carry with you most often? = I do not use a mobile phone*

**What is the highest level of education that you have completed?**
o   Less than a four-year degree and no more (such as high school or equivalent, some college, or a technical or associate's degree)
o   Four-year degree or higher (such as a bachelor's, professional, or graduate degree)

*Display This Question:*
*If What is the highest level of education that you have completed? = Less than a four-year degree and no more (such as high school or equivalent, some college, or a technical or associate's degree)*

**Currently, are you in school full-time to pursue a college degree?**
o   Yes
o   No

*Display This Question:*
*If What is the highest level of education that you have completed? = Four-year degree or higher (such as a bachelor's, professional, or graduate degree)*

**Do you have a doctorate (ex: PhD, EdD)?**
o   Yes
o   No

**In total, how much money did your household earn last year (2022)?**
o   Up to $50,000
o   $50,000 to $99,999
o   $100,000 or more

*Display This Question:*
*If In total, how much money did your household earn last year (2022)? = Up to $50,000*

**Did your household earn $26,500 or less last year (2022)?**
o   Yes
o   No

**Including yourself, how many people are in your household currently?**
________________________

**What is your gender?**
o   Male
o   Female
o   Non-binary or gender non-conforming
o   Prefer to self-describe ______________
o   Prefer not to say

**Are you Hispanic, Latino or Spanish?**
o   Yes
o   No
o   Prefer not to say

**What is your racial/ethnic identity?**
o   White or Caucasian
o   Black or African
o   Native American or Alaska Native
o   Asian - East or Central Asian
o   Asian - South, Southeast, or Southwest Asian
o   Native Hawaiian or Pacific Islander
o   Middle Eastern or North African
o   Prefer to self-describe ______________
o   Prefer not to say

**In the past week, how many hours did you use a mobile phone for any reason (such as sending or receiving calls, emails or text messages, taking or editing photos, playing games, or checking or updating social media accounts)?**
o   Less than 6 hours
o   6-10 hours
o   11-20 hours

- o  21-30 hours
- o  More than 30 hours

**Overall, how much experience have you had working with sensitive data (such as government data for which a security clearance is required, health data protected by HIPAA, or education data protected by FERPA)?**
- o  None at all
- o  Only a little
- o  A moderate amount
- o  A lot
- o  A great deal

**Which job category is the best match for your current situation?**
- o  Full-time employed outside the home
- o  Part-time employed outside the home
- o  Stay-at-home parent
- o  Student (full-time)
- o  Student (part-time)
- o  Unemployed
- o  Other _______________

**Which of the following best describes your current job?**

- o  Architecture and Engineering
- o  Arts, Design, Entertainment, Sports, and Media
- o  Building and Grounds Cleaning and Maintenance
- o  Business and Financial Operations
- o  Computer and Mathematical
- o  Community and Social Service
- o  Construction and Extraction
- o  Educational Instruction and Library
- o  Farming, Fishing, and Forestry
- o  Food Preparation and Serving Related
- o  Healthcare Practitioners and Technical
- o  Healthcare Support
- o  Installation, Maintenance, and Repair
- o  Legal
- o  Life, Physical, and Social Science
- o  Management
- o  Office and Administrative Support
- o  Personal Care and Service
- o  Production
- o  Protective Service
- o  Sales and Related
- o  Transportation and Material Moving

### 7.1.3 Directions for the "Yourself" Persona Assignment

For the following three screens, please rate the following: How likely is it that the given SMS text message is a legitimate message or a scam message, What would you be likely to do in response, and why. In each case, make your choice in terms of how you feel right now, not what you have felt in the past or would like to feel.

### 7.1.4 Directions for the "Pat Jones" Persona Assignment

For the following three screens, pretend that you are Pat Jones, who works at Baton Rouge University. Pat has many accounts with different banks, shopping websites, delivery services, and social media platforms. Pat's job makes it very important to respond promptly to legitimate text messages and to not respond to fraudulent text messages. For each message, please rate the following as if you were Pat: How likely is it that the given SMS text message is either a legitimate message or a scam message - What would you be likely to do in response, and why. In each case, make your choice in terms of how you feel right now, not what you have felt in the past or would like to feel.

### 7.1.5 Message-Specific Questions

**Remember to answer each question as <yourself/Pat Jones>.**

*<code to insert image>*

**Indicate the degree to which you (<Persona>) think this message is legitimate or a scam:**

- o  Almost certain to be a scam
- o  More likely to be a scam than to be legitimate
- o  Equally likely to be either legitimate or a scam
- o  More likely to be legitimate than to be a scam

o        Almost certain to be legitimate                    ☐        Other
                                                            _______________________________________
**If you (<*Persona*>) received this message on your
mobile phone, what would you do in response?**              **Why would you respond this way? Select all that
**Select all that apply.**                                  apply.**
☐        Reply by SMS text message to provide               ☐        Sense of urgency
information                                                 ☐        Curiosity
☐        Reply by SMS text message with STOP,               ☐        Trust in the sender
BLOCK, or other codes intended to stop receiving            ☐        Trust in the link URL (if included)
messages from the sender                                    ☐        Seeking a good outcome for myself
☐        Report the message using your device               ☐        Seeking to avoid a bad outcome for myself
options (such as clicking Block This Caller or Report       ☐        Lack of interest in the message
Junk)                                                       ☐        Other
☐        Click on the link in the SMS text message          _______________________________________
☐        Copy and paste the link into your phone's
web browser                                                 **To the best of your knowledge, do you,**
☐        Type the link into your phone's web browser        **<yourself/Pat Jones>, have an account with the**
☐        Forward the SMS text message to someone            **entity mentioned in the text message?**
else                                                        o        Yes
☐        Delete the SMS text message                        o        No
☐        Keep, save, or archive the SMS text message        o        Not sure
☐        Ignore the SMS text message

*7.1.6 Attention Check*

**We use this question to test whether survey participants are paying attention. Select the fourth option below (50%**
**to 74% of the time) to preserve your answers and continue on with the survey.**
o        Never
o        Up to 25% of the time
o        25% to 49% of the time
o        50% to 74% of the time
o        75% to 100% of the time

*Skip To: End of Block If We use this question to test whether survey participants are paying attention. Select the fourth...*
*!= 50% to 74% of the time*

*7.1.7 Security and Privacy Experiences*

**How frequently or infrequently have you personally**           o        Frequently
**been the victim of a breach of security (e.g. account**        o        Very frequently
**hacking, viruses, malware or theft of your personal**
**data)?**                                                       **To the best of your knowledge, how frequently or**
o        Very infrequently                                       **infrequently has someone close to you (e.g. spouse,**
o        Infrequently                                            **family member or close friend) been the victim of a**
o        Neither infrequently or frequently

**breach of security (e.g. account hacking, viruses, malware or theft of your personal data)?**
o    Very infrequently
o    Infrequently
o    Neither infrequently or frequently
o    Frequently
o    Very frequently

**How much have you heard or read about online security breaches?**
o    None at all
o    A little
o    A moderate amount
o    A lot
o    A great deal

**How much formal training have you had (either in school or on the job) in how to stay alert for security and privacy threats to your online data and accounts?**
o    None at all
o    A little
o    A moderate amount
o    A lot

o    A great deal

**Within the past three months, to the best of your knowledge, have you clicked on a fraudulent link in an email, text message, or web post?**
o    Yes, but it turned out to be a test being conducted as part of security awareness training
o    Yes, and it turned out to be a scam, but nothing bad happened, to my knowledge
o    Yes, and it turned out to be a scam, and I suffered a bad outcome (such as malware or theft of account credentials)
o    No, because I figured out that the links were fraudulent and actively avoided clicking on them
o    No, I have not noticed any fraudulent links in email, text messages, or web posts
o    I'm not sure

**Have you received security training that helps you identify fraudulent links or other threats in text messages?**
o    Yes
o    No
o    Not sure

*7.1.8 Security Attentiveness and Engagement – SA6/SA13 [14,15]*

Below is a series of statements about the use of security practices. Examples of security practices include using a password manager, using spam email reporting tools, installing software updates, using secure web browsers, activating biometric ID, and updating anti-virus software. For each, please indicate the degree to which you agree or disagree with each statement. In each case, make your choice in terms of how you feel right now, not what you have felt in the past or would like to feel. There are no wrong answers.

**I seek out opportunities to learn about security practices that are relevant to me.**
o    Strongly disagree
o    Disagree
o    Neither agree nor disagree
o    Agree
o    Strongly agree

**I am extremely motivated to take all the steps needed to keep my online data and accounts safe.**
o    Strongly disagree
o    Disagree
o    Neither agree nor disagree
o    Agree
o    Strongly agree

**Generally, I diligently follow a routine about security practices.**
o    Strongly disagree
o    Disagree
o    Neither agree nor disagree
o    Agree
o    Strongly agree

**I often am interested in articles about security threats.**
- o   Strongly disagree
- o   Disagree
- o   Neither agree nor disagree
- o   Agree
- o   Strongly agree

**I always pay attention to experts' advice about the steps I need to take to keep my online data and accounts safe.**
- o   Strongly disagree
- o   Disagree
- o   Neither agree nor disagree
- o   Agree
- o   Strongly agree

**I am extremely knowledgeable about all the steps needed to keep my online data and accounts safe.**
- o   Strongly disagree
- o   Disagree
- o   Neither agree nor disagree
- o   Agree
- o   Strongly agree

*Proactive Awareness – SeBIS [11]*

The following statements are meant to measure the frequency of actions taken to safeguard online security. Please read each statement and indicate the frequency with which you carry out or intend to carry out the action. There are no wrong answers.

**When someone sends me a link, I open it without first verifying where it goes.**
- o   Never
- o   Rarely
- o   Sometimes
- o   Often
- o   Always

**I know what website I'm visiting based on its look and feel, rather than by looking at the URL bar.**
- o   Never
- o   Rarely
- o   Sometimes
- o   Often
- o   Always

**I submit information to websites without first verifying that it will be sent securely (such as by checking for SSL, "https://", a lock icon).**
- o   Never
- o   Rarely
- o   Sometimes
- o   Often
- o   Always

**When browsing websites, I mouseover links to see where they go, before clicking them.**
- o   Never
- o   Rarely
- o   Sometimes
- o   Often
- o   Always

**If I discover a security problem, I continue what I was doing because I assume someone else will fix it.**
- o   Never
- o   Rarely
- o   Sometimes
- o   Often
- o   Always

### 7.1.9 Social Strategy – SSBS [19]

**I care about the source of the mobile app when performing financial and/or shopping tasks on that app.**
o Never
o Rarely
o Sometimes
o Often
o Always

**When downloading a mobile app, I check that the app is from the official/expected source.**
o Never
o Rarely
o Sometimes
o Often
o Always

**Before downloading a mobile app, I check that the download is from an official application store.**
o Never
o Rarely
o Sometimes
o Often
o Always

**I verify the recipient/sender before sharing text messages or other information using mobile apps.**
o Never
o Rarely
o Sometimes
o Often
o Always

**I delete any online communications (i.e., texts, emails, social media posts) that look suspicious.**
o Never
o Rarely
o Sometimes
o Often
o Always

**I pay attention to the pop-ups on my mobile phone when connecting it to another device (e.g. laptop, desktop).**
o Never
o Rarely
o Sometimes
o Often
o Always

**I pay attention to the pop-ups that show up on ANY of my devices (e.g. laptop, desktop, tablet, mobile phone).**
o Never
o Rarely
o Sometimes
o Often
o Always

### A.2  Open-ended Responses to Messages

Table 8: A list of the text responses entered with the "Other" choice for items asking how the participant would respond to the given text messages, and why they would respond that way.

|    | How to respond | Why that response |
| --- | --- | --- |
| R1 | Contact school tech team to see if this is a true message sent by them.<br>report to phone service provider<br>Contact BRU<br>block number<br>Check for a web site for BRU.edu<br>Links can be manipulated. <a href="xxx.html">aaa.html</a> | No<br>If it is a scam, my personal data will be compromised.<br>Caution<br>I would have known about this in advance. I wouldn't be texted out of the blue.<br>I usually don't click prize scams in SMS messages. Too much of a security issue, unless if it's from the company/business's official number/account. |
| F1 | report to phone service provider<br>Action dependent if we had an account with Walmart | To not fall for the scam<br>I won't respond back. |

|    | How to respond | Why that response |
|----|----------------|-------------------|
|    | Ignore the text unless I felt that I wanted to take the survey... just like I am taking this one.<br>Call walmart<br>report to walmart | Scam<br>the return web address is not walmarts web address. |
| R2 | Check my bank account<br>Look into my bank account and see if there was a transaction made that I did not do<br>Google for similar text messages<br>Call chase<br>Call the credit card company directly<br>Call company<br>I would call Chase to see if they sent the text<br>[C]all the bank<br>Go to my bank website directly without using a link to check and see if there are notifications on my account page. Also google to see if similar texts have been reported as scams.<br>I would contact a legitimate number from my info when opening my account to inquire about any fraudulent activeit on my accounty<br>depends on if I made a purchase<br>If I had a Chase card with those ending digits, I'd call their customer support directly to resolve the issue.<br>go directly to chase and see if that amount was charged.<br>If I actually have a Chase card and actually did use it for gas on that date, then I would respond to the msg. Otherwise, I would report it to Chase as potential fraud.<br>Log on to my Chase account and react to alert, there.<br>Call bank on phone number I know is correct<br>I would not reply to anything, but I would call Chase bank to see if this is a scam<br>If I had a card with them I would call and check it out<br>Call chase<br>Telephone bank cust svc number to discuss<br>Phone Chase Bank to verify<br>Check my Chase account online<br>Forward the message to my bank. Banks don't put RATES MAY APPLY in the message. | I wouldnt<br>Knowing I'd probably get my information stolen<br>Safer than clicking on fraudulent link or responding to text<br>Don't like scams<br>I am proactive.<br>Prefer to work directly on my bank site, for all inquiries and transactions<br>Link begins with HTTPS<br>Think it's a scam<br>Delete with no response<br>I don't click links unless family. |
| F2 | would call them<br>check the status of my debit card on the providers website<br>I don't know<br>type it into a web browser in a separate device<br>Call the bank<br>Call the company<br>call my bank<br>call bank of america<br>Call my credit card company to verify this is true information<br>Sign into my account and see if it works.<br>Call bank to check if legitimate<br>call my bank<br>If I actually had a BOA account, I would contact them directly (not via a reply to the SMS) and verify the info. If it is legit, I would take appropriate action(s).<br>I would call bank of America to let them know I received this message<br>Contact the bank by phone to verify they sent the text.<br>I'd contact the number for customer service on the back of my card and check with a real person<br>call the bank directly<br>Forward it to whoever handles phishing<br>If I did have an account with them i would alternately verify that it was indeed blocked and and then remedy.<br>Telephone my bank cust svc number to ask if they sent msg<br>call bank of America | leave it alone<br>I don't know<br>I won't respond back nor click on the link.<br>Just doesn't look real<br>probable scam<br>the senders email address is not consistant with a address from BOAt<br>I want to see if this is a scam<br>Don't know how to do computer stuff on phone. |

| | How to respond | Why that response |
|---|---|---|
| R3 | (no Other responses) | scam<br>think its a scam and don't want to risk it.<br>Not relevant to me in any way<br>scam |
| F3 | Also, block the number<br>Delete<br>Would report to amazon, who does follow up on scammers using their brand name. | No<br>possible scam<br>Dont need/want a job<br>Red flags in this. |
| R4 | Call company<br>Spam at least, to dangerous to click link for any reason DELETE | I seek out what I want and don't respond to things I didn't initiate<br>I won't respond back.<br>scam<br>am not near bankruptcy<br>I don't believe that they would send a message<br>Unknown individual sending me an interested text<br>SPAM |
| F4 | block number | I don't click on random links sent to me regardless of their claim<br>to good to be true usually means.............<br>Report scammers to provider<br>Scam<br>Caution<br>Looks like a scam<br>spam at the least, dangerous to click link DELETE |
| R5 | Text ginger and ask if she sent it or say something she would respond to to verify its her<br>If I really went to a birthday party I'd click the link<br>Message back to talk to someone to make sure it is reL<br>Select the link only if I had previous knowledge of what the message was about<br>I would try to remember if there is a Ginger among my friends<br>respond if i know ginger<br>Call ginger yo see if she really sent the text<br>text the person who had the party to see if they sent it to me.<br>Don't know how to read texts.<br>If we knew Ginger, and M and attended the birthday party, we would respond. | No<br>Scammers need to be reported<br>Don't read or answer texts. |
| F5 | (no Other responses) | scam<br>NOT A LEGIT EMAIL ADDRESS<br>Scam. I see this a lot |
| R6 | wait for the next sms message.<br>Try remember text<br>Read it, possibly save it, or delete it after<br>Keep a look out for the vehicle<br>nothing unless I have info then I would call 911 to give the info I had<br>Keep a lookout for the vehicle.<br>Turn on the radio or the news<br>Review it and be on lookout for said vehicle. Later on would delete.<br>Be on the lookout for that license #<br>Nothing much; would read, look up news, and clear emergency alerts in my storage. My notification repeats until disabled.<br>Ignore it unless I spotted a Tan Honda Civic with the license plate listed in the text.<br>Call Law enforcement to see if it is valid. | moral obligation to keep children safe<br>An amber alert wouldn't have a link to push<br>I would believe it but wouldn't click anything<br>there's no description of the kidnapped person in the message at all, so it's most likely a fraudulent text<br>keep alert from that txt, but nothing further of action to be done.<br>Right thing to do<br>Amber alerts don't have links<br>Has a .gov address so will be on alert for car but won't click the link<br>Important to be on lookout for missing children.<br>would mostly like not know anything to reply to message<br>Didn't delete incase I would happen to see something.<br>No need to respond unless I see the vehicle in question.<br>no knowledge of it<br>Amber alerts come as a notification, not as a text<br>Keep myself safe while being a good citizen<br>Skepticism<br>I generally don't click on links unless sent from personal contact<br>By calling 911 it would validate if the message was a scam<br>I would save it to call in case I saw the vehicle in question. |

| | How to respond | Why that response |
|---|---|---|
| | | Not living near that state<br>look out for a Honda with Texas tags.<br>Amber alerts do not come over sms texts |
| F6 | copy it into a browser on a separate device as an anonymous report to phone service proivider<br>look up the phone number<br>the IRS only send a notice in the mail | To be safe<br>IRS communicates by certified mail.<br>SPAM<br>Because the government wouldn't send a text like that<br>I don't respond to spam<br>I feel like all the messages are scam nowadays<br>The IRS only notify you by mail<br>It's a scam<br>Caution<br>Highly suspect message<br>IRS only contacts people thru the USPS<br>IRS would not send a text<br>It's wouldn't send a message.<br>I know the IRS would never send me a text message |
| R7 | log on into Amazon's account<br>Forward to Amazon<br>Go to my Amazon account<br>Block sender<br>I'd just get on my Amazon app<br>Go to my Amazon account and look for notifications<br>Check my account<br>Go to the Amazon web site directly to see what's up<br>Block number<br>Use my internet to contact Amazon to see if the message is real or a fruad.<br>contact Amazon directly, not use the text<br>Report to Amazon<br>call Amazon to verify<br>Check email or go to Amazon site<br>Need to tell me where my package is to show me if there is a problem. | it will be accurate<br>My shipping's contact info, where "262966" is in image, is saved from initial app verification code sent. Any other's are ignored. I ignore things like this and go to app or website. And Gmail; I check sender's address there<br>it would depend on if I had ordered anything from Amazon<br>Needs to be from the official carrier to be legitimate. |
| F7 | Check to see if my FB account is actually suspended and if it hasn't, I would report the sender.<br>I would check my fb, but not from that link<br>Sign into Facebook to see if there are any problems. If necessary send a message within Facebook to ask if there is a problem.<br>check my facebook account | Safer to verify the information in another way rather than responding to possible fraudulent texts.<br>Hate scams and spam<br>Scam<br>I don't use social Media<br>Scam |